\documentclass[aps,pre,twocolumn,groupedaddress]{revtex4-1}
\usepackage{amsmath}
\usepackage{graphicx}
\usepackage{caption}
\usepackage{subcaption}

\begin{document}

%\preprint{}

\title{Geometrical model for malaria parasite migration in structured environments}


\author{Anna Battista$^{1,2}$, Friedrich Frischknecht$^3$, Ulrich S.~Schwarz$^{1,2}$}
\email[]{ulrich.schwarz@bioquant.uni-heidelberg.de}
%\thanks{}
\affiliation{$^1$Institute for Theoretical Physics, Heidelberg University, Heidelberg, Germany}
\affiliation{$^2$BioQuant, Heidelberg University, Heidelberg, Germany}
\affiliation{$^3$Department of Infectious Diseases, University Clinics Heidelberg, Heidelberg, Germany}


\date{\today}

\begin{abstract}
Malaria is transmitted to vertebrates via a mosquito bite, during which rod-like
and crescent-shaped parasites, called sporozoites, are injected into the skin of the host. 
Searching for a blood capillary to penetrate, sporozoites move quickly in locally helical trajectories,
that are frequently perturbed by interactions with the extracellular environment. 
Here we present a theoretical analysis of the active motility of sporozoites in a structured environment.
The sporozoite is modelled as a self-propelled rod with spontaneous curvature and bending rigidity.
It interacts with hard obstacles through collision rules inferred from experimental observation of
two-dimensional sporozoite movement in pillar arrays.
Our model shows that complex motion patterns arise from the geometrical shape of the parasite and
that its mechanical flexibility is crucial for stable migration patterns. 
Extending the model to three dimensions reveals that a bent and twisted rod can
associate to cylindrical obstacles in a manner reminiscent of the association of sporozoites to blood capillaries,
supporting the notion of a prominent role of cell shape during malaria transmission.
\end{abstract}

%\keywords{}

\maketitle

\section{Introduction \label{sec:intro}}
Malaria is caused in vertebrate hosts, for example mice and men, by eukaryotic parasites belonging to the genus \textit{Plasmodium}. The parasites are transmitted to the vertebrate host via the bite of an infected \textit{Anopheles} mosquito. During the blood meal, the mosquito injects several tens of parasites, called \textit{Plasmodium sporozoites}, into the skin of the host. A first necessary condition for the host to become infected is that some of these sporozoites reach and penetrate a blood vessel \cite{Sidjanski1997,Vanderberg2004,Amino2006,Menard2013}, to be then transported to the liver by the blood stream. After a replication stage in the liver, the parasites enter again the blood circulation and infect red blood cells, leading to the typical symptoms of the disease, like periodic fevers. The infectious cycle continues with the uptake of infected blood by another mosquito.

Although each stage of the malaria life cycle offers interesting physical aspects, here we focus on sporozoite migration because it is characterised by some surprising features that have not been the object of a mathematical analysis before. We first note that sporozoite migration in the skin is very fast, with an average migration speed exceeding $1\ \mu m/s$, which is 1-2 orders of magnitude faster than the migration of typical tissue cells \cite{Amino2006}.
Plasmodium sporozoites are about $10\ \mu m$ long and about $1\ \mu m$ wide, depending on the species and stage of maturity \cite{Kudryashev2012}. Mature sporozoites are crescent shaped with an average radius of curvature of about $5\ \mu m$ \cite{Vanderberg1974} and move on circular trajectories on a flat substrate, with a strong preference for counterclockwise movement (under an inverted microscope, that mirrors the real image) \cite{Vanderberg1974}. 
In contrast to crawling cells, that extend protrusions to translocate, sporozoites migrate with changes
in cellular shape that are limited to their curvature \cite{Munter2009}. 
Sporozoites lack cilia or flagella and therefore are not swimming cells, as for example \textit{E.~coli} \cite{Lauga2006}. Instead, their type of motion is called \textit{gliding motility} and depends on adhesion between the cell and the substrate. 
Sporozoites are supposedly propelled by an internal conveyer belt system that translocates adhesive transmembrane proteins from the front to the rear end of the cell, exploiting an evolutionary ancient actomyosin system \cite{Baum2006,Montagna2012}.
In contrast to mammalian tissue cells, however, this adhesion is relatively non-specific as the adhesion receptors seem to bind to many different ligands \cite{Perschmann2011}.

The observation of 2D circular movement, together with the fact that, in an unstructured 3D environment, sporozoites tend to move on roughly helical paths, suggest that the movement of sporozoites is strongly determined by their geometrical shape \cite{Vanderberg1974,Amino2006,Amino2008,Hellmann2011,Kudryashev2012,Akaki2005}. However, sporozoite trajectories are also highly dependent on the structural properties of their environment. For example, they show different patterns of locomotion in
the tail versus the ear of mice, possibly as a result of the different organisation of the extracellular matrix \cite{Hellmann2011}.
The role of the environment in sporozoite migration has been previously addressed by experimental studies that have employed micro-fabricated pillar arrays \cite{Hellmann2011}. Similar experimental approaches have been used recently also for other cellular systems, for example the nematode worm \textit{C.~elegans} \cite{Johari2013}, the amoeba \textit{Dictyostelium} \cite{Arcizet2012}, the bacteria \textit{Neisseria gonorrhoeae} and \textit{Myxococcus xanthus} \cite{Meel2012} and the eukaryotic parasite causing the sleeping disease, the trypanosome \cite{Heddergott2012}.
Migration of sporozoites through arrays with different pillar diameters and spacing revealed complex motion patterns depending on both radius and spacing, highlighting the role of the interplay between the peculiar geometrical features of the parasite and repeated encounters with obstacles \cite{Hellmann2011}. However, a quantitative framework to understand the features of sporozoite migration in complex environments is still missing. 

Sporozoite gliding is different from most other types of cell locomotion that have been analysed with mathematical models before, but also shares some instructive similarities. The flow of actin and related molecules driving the conveyer belt system \cite{Heintzelmann2006,Montagna2012} is reminiscent of the flow of actin in keratocytes, that has been studied e.g.~within a hydrodynamic framework \cite{Rubinstein2009} or phase field models \cite{Ziebert2013}.
However, the structural organisation of the actin system in sporozoites is very different and far from understood \cite{Schmitz2010}.
The coupling between the sporozoite and the substrate via adhesion molecules is similar to the situation in adherent animal cells, which has been described before e.g.~as a slip boundary for an active gel \cite{Joanny2003,Carlsson2011}, but again the exact details are not known yet. In particular it is unclear how adhesion molecules are injected into and removed from the conveyer belt system, and how they are organised along the body of the sporozoite \cite{Munter2009}.

Due to the uncertainty regarding the molecular system, it is natural to start with a mathematical analysis at the cellular level. 
In this framework, active motion in the presence of hard obstacles is often treated using Brownian models for simply shaped and non-deformable particles \cite{Peruani2007,VanTeeffelen2008,Reichhardt2013,Kuemmel2013}. However, in the case of sporozoites, as well as of other cells moving in structured environments \cite{Johari2013,Arcizet2012,Meel2012,Heddergott2012}, cell deformations should be taken into account. 
One instructive example is the bacterium \textit{Myxococcus xanthus}, that is also self-propelled and rod-shaped
and that has been described as a flexible chain of beads \cite{Starruß2007,Wu2007}. While this work investigated
interactions between the bacteria, here we address a single but curved object interacting with a given obstacle array.
Moreover we do not model the parasite as a chain of beads, but introduce a parametrisation that is computationally less expensive.

In this paper we propose a cell-level, geometry-based model for sporozoite migration that takes into account the sporozoite curvature and length, but not its thickness. Our model focuses on adhesion to a solid support and repeated collisions with obstacles. We exclude the explicit modelling of the internal gliding machinery whose effect, to a first approximation, is to propel the sporozoite forward at a typical speed. Instead, we model the parasite as a self-propelled rod with spontaneous curvature that, in the absence of obstacles, moves with constant speed along a circular trajectory of matching curvature.
Motivated by experimental observations, collisions with hard obstacles are solved either by a temporary change
in the curvature of the rod or by a random reorientation. 
Extending our model to three dimensions, we consider the effect of twist and show that both bending and twisting help the parasite to navigate arrays of cylindrical obstacles with the shape of blood vessels.
Our analysis demonstrates how the complexity of sporozoite motility in structured environment is shaped by the geometrical and mechanical properties of the parasite, and provides a first quantitative framework to rationalise the experimental results for sporozoites moving in pillar arrays and other structured environments.

\section{Experimental observations \label{sec:exp_motiv}}

\begin{figure*}
 \includegraphics{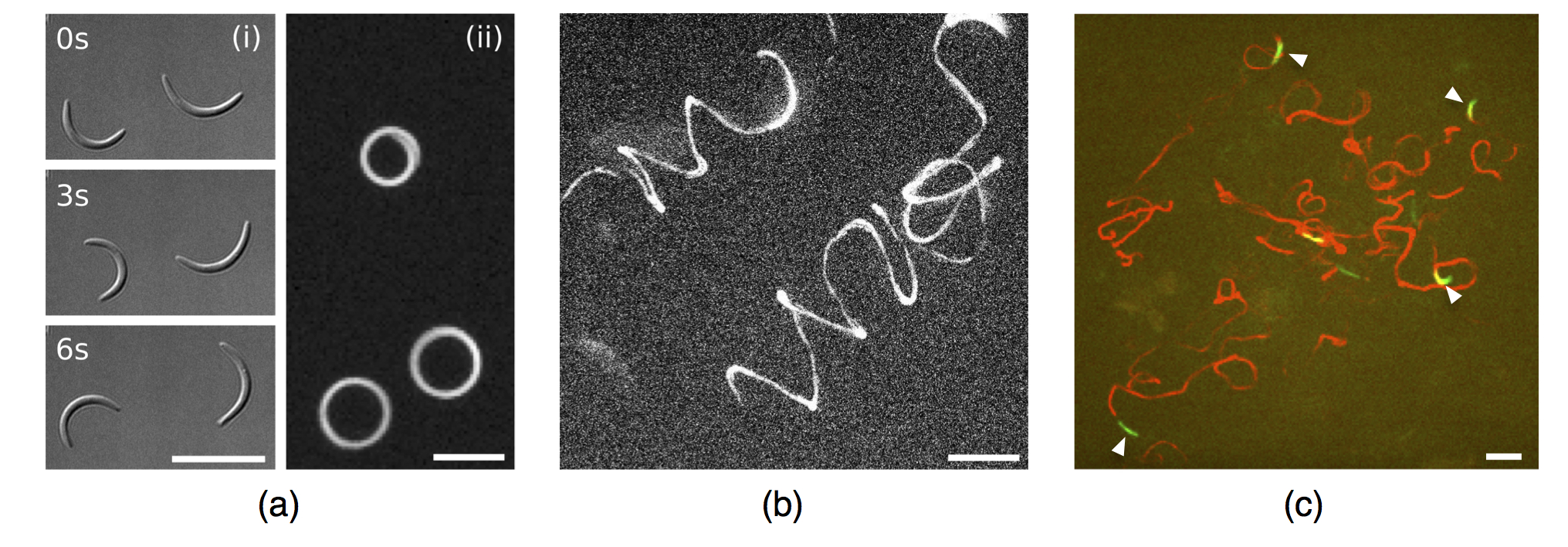}
 \caption{\label{fig:nonstruct} Plasmodium sporozoites are curved gliding parasites that describe environment-dependent trajectories.
(a) On a 2D substrate sporozoites move on roughly circular trajectories in a counterclockwise direction. (i) Time series of differential interference contrast (DIC)
images (time given in seconds). (ii) Overlay of a time series of fluorescent images of three gliding sporozoites.
(b) In 3D unstructured gels sporozoites describe perturbed helical trajectories, as shown here with maximum fluorescent projections of two sporozoites migrating in matrigel. (c) \textit{In vivo} trajectories (red) of sporozoites (green, white arrowheads) migrating in the dermis of the ear of a mouse appear more random but circular elements are clearly present. The trajectories are maximum fluorescence intensity projections for 240s. Sporozoites are represented at the start of the imaging in green and pointed to by the white arrowheads. Trajectories, in red, are represented from 5s after the start till the end of the imaging time. All scale bars: $10\ \mu m$.
}
\end{figure*}

We start with the experimental observations underlying our theoretical analysis
(more details of our experimental procedures are provided in appendix A). 
Sporozoites from the rodent model parasite \textit{Plasmodium berghei} migrating on a two
dimensional flat substrate describe circular trajectories with a radius similar to
their own radius of curvature (fig.~\ref{fig:nonstruct}a). Migration in unstructured 3D environments,
such as matrigel, leads to perturbed helical trajectories (fig.~\ref{fig:nonstruct}b). Trajectories \textit{in vivo}
(e.g.~in the ear of a mouse) are more irregular, but still include circular elements with a radius that again seems to correspond
to the radius of curvature of the parasite (fig.~\ref{fig:nonstruct}c). 

\begin{figure}
 \includegraphics{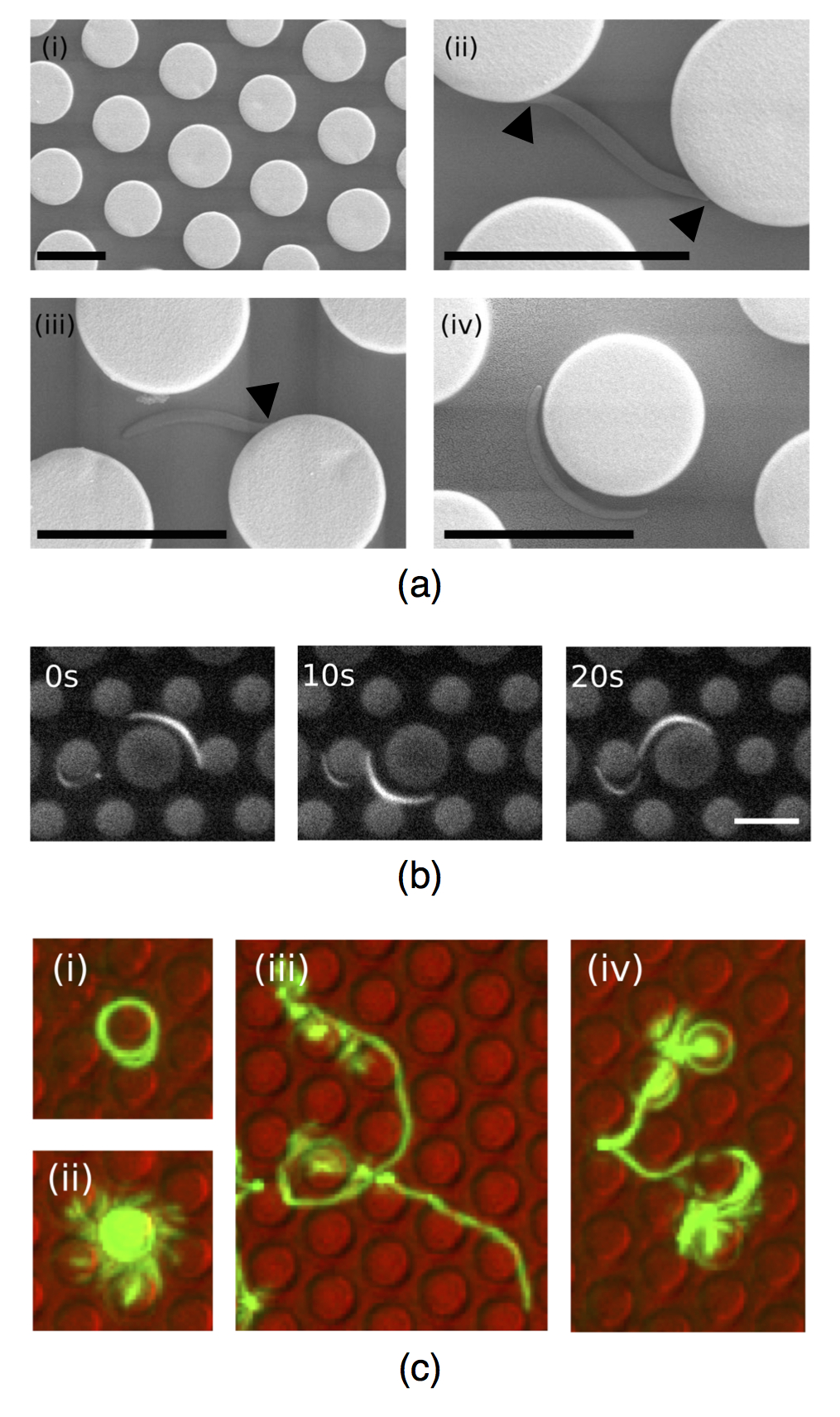}
 \caption{\label{fig:struct} (a) (i) Top views by scanning electron microscopy of micro-fabricated arrays consisting of PDMS pillars on a flat PDMS substrate and arranged according to a hexagonal lattice. (ii-iv) Plasmodium sporozoites migrating in the arrays lie on the substrate and can interact with the pillars: the ends of the parasite can either contact the pillars (black arrowheads) or not. 
 %See also Movie S1.
 Scale bars: $10\ \mu m$. (b) Circling of sporozoites around pillars does not require direct contact. Furthermore, occasional contact does not confine the parasite to the pillar. 
 %See also Movie S2.
 Numbers indicate time in seconds. Scale bars: $10\ \mu m$. (c) Typical trajectories of sporozoites migrating in pillar arrays: circular trajectories (green) around pillars (red) (i), linear (iii) and meandering (iv) trajectories. In addition, adhesion of sporozoites
can also be unstable, leading to a probing phase that ends with migration into a new direction (ii).}
\end{figure}

In order to study motility patterns in quantitative detail, micro-fabricated PDMS pillar arrays have emerged
as a very useful assay because it allows us to design the geometrical structure and to perform microscopy
that can be easily combined with image processing. The pillars are usually placed in a hexagonal arrangement (fig.~\ref{fig:struct}a(i)).
Because sporozoite adhesion is relatively non-specific, no special surface functionalization or passivation is required.
The sporozoites settle between and interact with the pillars, sometimes leading
to clear deviations from their unperturbed shape (fig.~\ref{fig:struct}a(ii)).
While sporozoites often seem to make contact with a pillar during translocation (fig.~\ref{fig:struct}a),
adhesion between the parasite and a pillar is not a requirement to circle around it (fig.~\ref{fig:struct}b).
Typical trajectories of P.~\textit{berghei} sporozoites in pillar arrays are shown in fig.~\ref{fig:struct}c as 
overlay of a time series of fluorescent images. Sporozoites can circle around pillars both in a stable manner (fig.~\ref{fig:struct}c(i))
and in an unstable manner (fig.~\ref{fig:struct}c(ii)). In the second case, the parasite remains attached to the substrate with its back end and probes the environment in a waving manner until it adheres and migrates again with a new orientation.
Trajectories can also show distinct linear patterns as in fig.~\ref{fig:struct}c(iii) or be mainly meandering as in fig.~\ref{fig:struct}c(iv).

Although motility patterns of sporozoites in obstacle arrays can be automatically classified using image
processing \cite{Hegge2009,Hellmann2011}, it is difficult to quantify the systematic effect of the structure of the environment
on the migration of the single cell because of the inherent variability within a population of sporozoites.
Experimental evidence suggests that spontaneous curvature, cell length, bending rigidity and adhesion
structure show a strong degree of variability from one sporozoite to the other. For example, fig.~\ref{fig:hist_curv}
shows the broad distribution of spontaneous curvature (details given in appendix A).
While this geometrical feature of the sporozoite can be measured
relatively easily, most of its other features are not directly accessible, especially not in the context of motility assays.
In the present work we propose a physical model to systematically study the role of the geometrical and mechanical properties
of sporozoites during migration in obstacle arrays, which in the future might be helpful to estimate these
microscopic properties from macroscopic observations. 
\begin{figure}
 \includegraphics{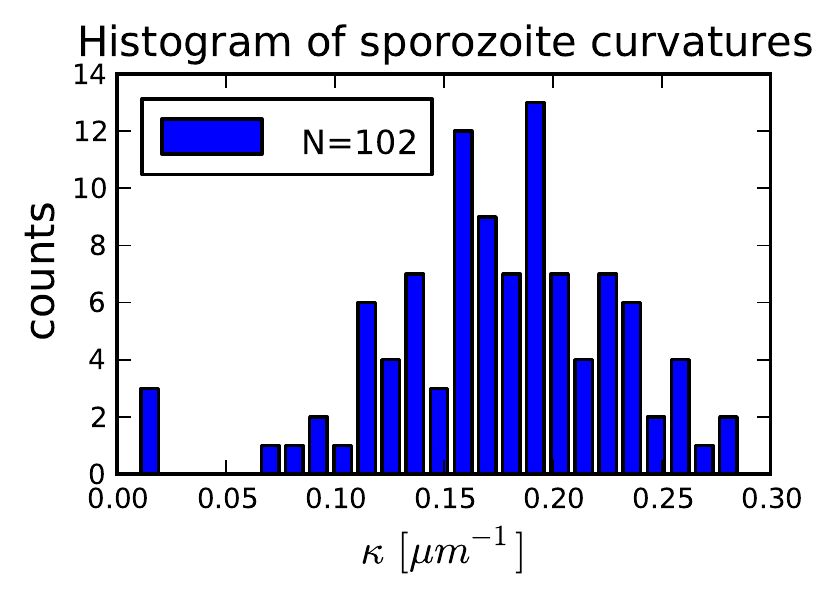}
 \caption{\label{fig:hist_curv} Curvature distribution of Plasmodium sporozoites automatically measured from a sample of $N=102$ cells. The distribution is widely distributed around curvatures of about $0.2\ \mu m^{-1}$, corresponding to an average radius of curvature of about $5\ \mu m$.}
\end{figure}

\section{Basic model \label{sec:model}}

\subsection{Geometrical considerations \label{sec:geometry}}

\begin{figure}
 \includegraphics{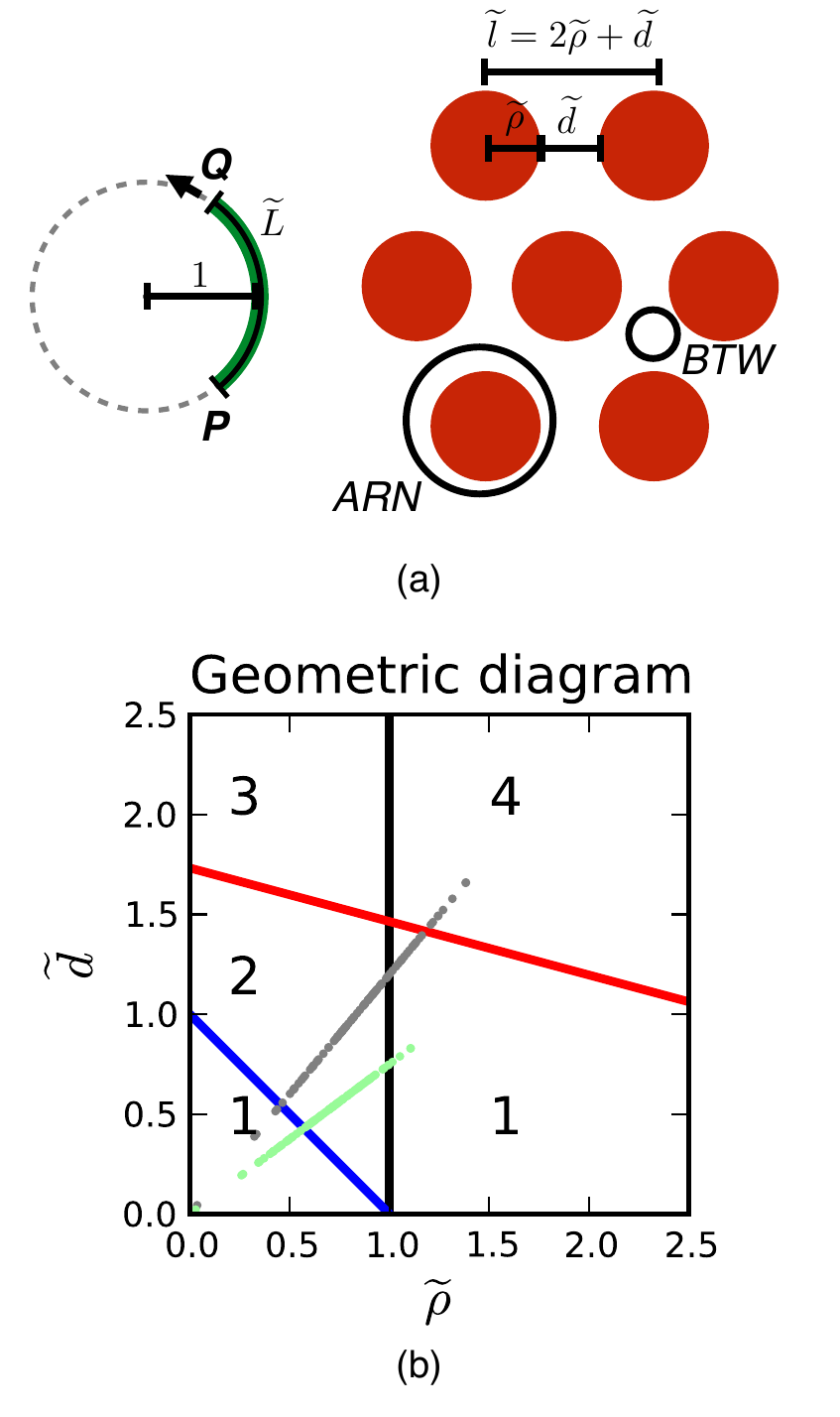}
 \caption{\label{fig:system_diagr} (a) Plasmodium sporozoites are elongated cells, with an aspect ratio of about 10. We consider a 1D model parasite as a bent rod that, when unperturbed, describes a circular trajectory on the circle that best fits its shape. Migration of the rod in hexagonal obstacle arrays is characterised by frequent collisions, but two unperturbed stable motility patterns are possible, depending on the obstacle size and distance: circling around one or more obstacles ($ARN$) and circling between the obstacles ($BTW$). (b) Geometric diagram identifying the regions of the $(\widetilde{\rho},\widetilde{d})$ plane, delimited by the solid lines, where the stable motility patterns $ARN$ and $BTW$ are allowed. In region 1 no steady pattern is possible, region 2 allows for $ARN$ patterns, region 3 for both $ARN$ and $BTW$ patterns, while in region 4 only $BTW$ patterns are possible. The dots correspond to the curvature values from the distribution of fig.~\ref{fig:hist_curv} for typical obstacle array parameters $(\rho=5,d=6)\mu m$ (gray) and $(\rho=4,d=3)\mu m$ (green) used in experiments \cite{Hellmann2011}.}
\end{figure}

Consider a hexagonal lattice of lattice constant $l$ patterned with circles (pillars) of radius $\rho$, the distance between two nearest neighbours being $d=l-2 \rho$. We model the sporozoite as a bent rod of length $L$ and curvature $\kappa_0$ (fig.~\ref{fig:system_diagr}a). We define a set of non-dimensional quantities such that the rod has unit curvature:
\begin{equation}
 \widetilde{\rho}=\rho\kappa_0;\enspace\widetilde{d} = d\kappa_0;\enspace\widetilde{l} = 2\widetilde{\rho}+\widetilde{d};\enspace\widetilde{L} = L\kappa_0\enspace.
\end{equation}
Consider the parasite moving on the circle that best fits its shape. In a $(\widetilde{\rho},\widetilde{d})$ lattice the parasite is said to move steadily if its circular trajectory does not collide with any of the obstacles in the lattice. This is possible in two ways: either the trajectory lies between the pillars ($BTW$) or it loops around one or more pillars ($ARN$) (fig.~\ref{fig:system_diagr}a). Conditions for steady movement in the obstacle array are derived from the geometry of the lattice. Circling between the pillars is possible for distances 
\begin{equation}
 \widetilde{d} \geq \sqrt{3} - \widetilde{\rho}(2-\sqrt{3}) \label{eq:d_btw}
\end{equation}
while circling around one or more pillars is possible for 
\begin{equation}
 \widetilde{d} \geq 1-\widetilde{\rho}\enspace\text{and}\enspace\widetilde{\rho}\leq1\enspace. \label{eq:d_arn}
\end{equation}
The curves identified above define four regions in the $(\widetilde{\rho},\widetilde{d})$ plane, delimited by the solid lines in fig.~\ref{fig:system_diagr}b. In region 1 no steady pattern is possible, region 2 allows for $ARN$ patterns, region 3 for both $ARN$ and $BTW$ patterns, while in region 4 only $BTW$ patterns can fit. The dark gray and green (or light grey) points  in fig.~\ref{fig:system_diagr}b correspond to the curvature values from the distribution of fig.~\ref{fig:hist_curv} for typical obstacle array parameters $(\rho=5\mu m,d=6\mu m)$ and $(\rho=4\mu m,d=3\mu m)$, respectively, as used in experiments \cite{Hellmann2011}. Since a sample of sporozoites does not localise
to a unique region of the $(\widetilde{\rho},\widetilde{d})$ plane, the information on the curvature of the sporozoite is necessary to make predictions about its most likely trajectory. For a typical experiment with the curvature distribution shown in fig.~\ref{fig:hist_curv}, we conclude
that both ARN and BTW motility patterns are expected to be observed.

\subsection{Interaction with the obstacles \label{sec:interaction}}

\begin{figure*}
 \includegraphics{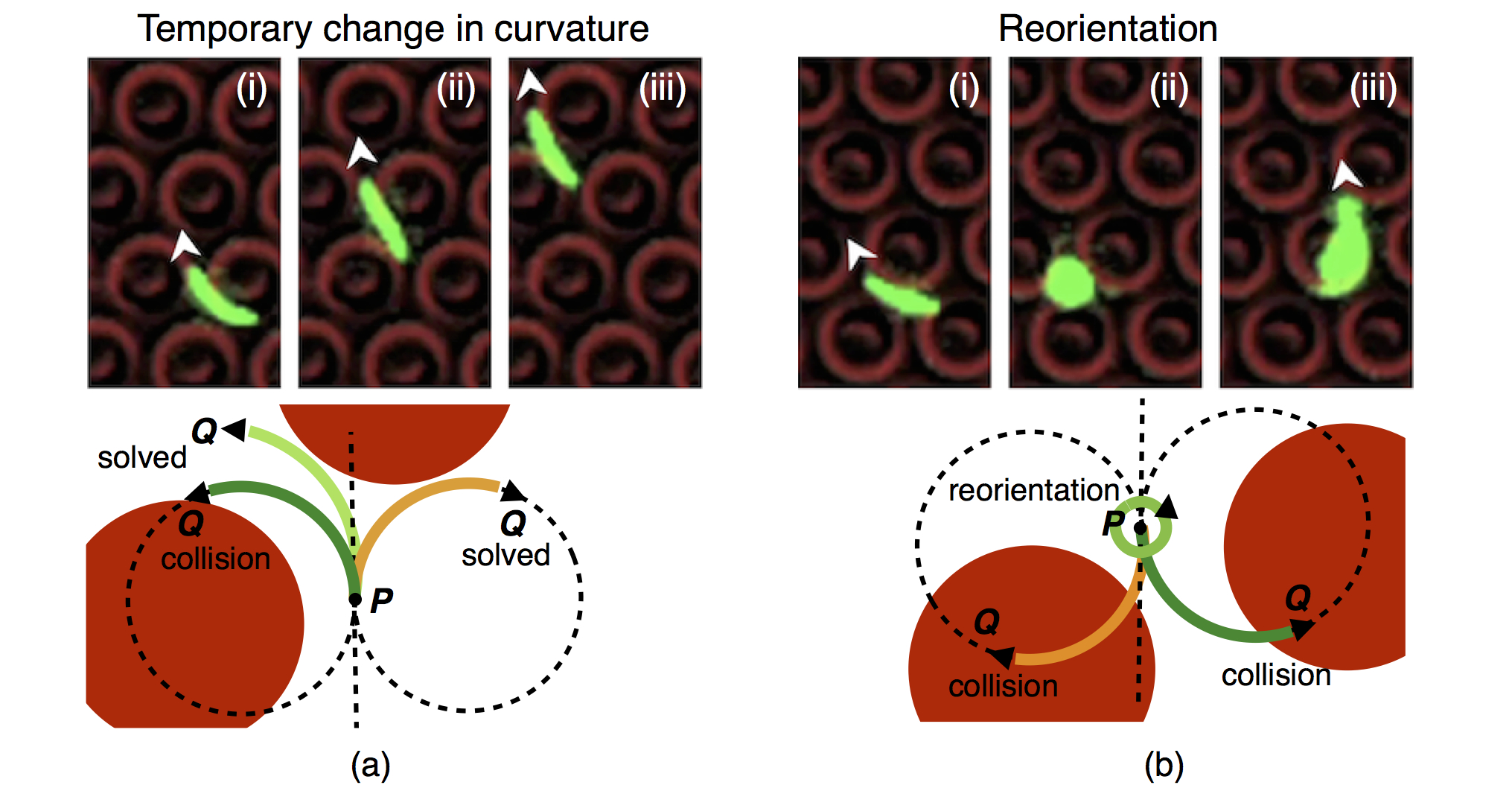}
 \caption{\label{fig:interaction} We group the interactions between Plasmodium sporozoites and obstacles in two categories. In the first (a), obstacles are avoided via a change in the bending of the cell body. The deformation is temporary, meaning that once the collision is solved the rod proceeds its migration with unit curvature, the sign of which is consistent with that of the temporary curvature, while the deformed region progressively shifts towards the rear of the rod. (b) In the second interaction mode the collision is solved via a reorientation of the rod, after which migration is resumed with unit curvature and an equal probability of moving clockwise or counterclockwise. Our model thus allows both for changes in the curvature of the rod as well as for random reorientations. The balance between the two modes of interaction depends on the model parameters.}
\end{figure*}

Motivated by detailed experimental observations of \textit{Plasmodium berghei} sporozoites migrating in pillar arrays,
we divide the interaction with the obstacles in two categories (fig.~\ref{fig:interaction}). In the first case, the parasite successfully avoids the pillar via a temporary change in the bending of the cell body. In the second case, the parasite transiently looses adhesion to the substrate to then reattach with a new orientation to resume migration. In the following we formalize these experimental observations using two different descriptions of the interaction between the bent rod and the obstacles: the binary and the statistical interaction schemes, corresponding to a stiff and a flexible rod, respectively. To keep the model simple, we
disregard the asymmetry between clockwise and counterclockwise motion, which is expected to be of minor relevance for the case of frequent collisions studied here.

In the binary model the sporozoite is stiff and can move with fixed unit curvature either clockwise or counterclockwise. A collision triggers a check to determine if the configuration symmetric with respect to the tangent in $P$ solves the overlap (fig.~\ref{fig:interaction}a, rightmost orange rod). If not, than the parasite reorients by rounding up around $P$ and sampling a random exit angle to resume motion with unit curvature and a probability $1/2$ of moving either clockwise or counterclockwise (fig.~\ref{fig:interaction}b). The binary model is purely geometrical and avoids changes in curvature, but allows for the two basic responses observed in experiments (deflection and tumbling).

In the statistical model, we allow the parasite to assume a non-natural curvature for a short time in order to solve a collision situation.
We consider an interval $I=[-\widetilde{\kappa},\widetilde{\kappa}]$ of curvatures that the sporozoite can assume in order to avoid an obstacle upon collision. However, only a subset $I_c\subseteq I$ allows the parasite to avoid the obstacle.
We assign to each curvature an energy $E(\kappa)$ according to a symmetric double well potential with minima in $\kappa=\pm 1$, so that in general the total bending energy of the rod reads
\begin{align}
 E_{bend} & = \int_0^{\widetilde{L}} ds\,E(\kappa(s)) = \nonumber \\ 
 & = \int_0^{\widetilde{L}} ds\left[ (\kappa(s)-1)(\kappa(s)+1) \right]^2 \enspace.\label{eq:en_bend}
\end{align}
We introduce a temperature via $\beta=1/(k_B T)$ to account for the strength of the bending effect. 
We also assign an energy $E_{tmb}$ to a reorientation (tumbling) event. We finally distinguish between
new curvatures of equal or of different sign. Therefore we introduce $\sigma$, the initial sign of the
curvature. We then define the partition function $Z(\beta)$ that includes all possible outcomes of the collision event
in the statistical model:
\begin{align}
 Z(\beta) = &\, e^{-\beta E_{tmb}} + \frac{1}{2}(1+\sigma\,\text{sgn}(\kappa)) \sum_{\kappa\in I_c}e^{-\beta E_{bend}} + \nonumber \\ 
 & + \frac{1}{2}(1-\sigma\,\text{sgn}(\kappa)) \sum_{\kappa\in I_c}e^{-\beta E_{max}} \enspace. \label{eq:Zc} 
\end{align}
This equation shows that the rod has three possibilities upon collision: it can tumble, it can assume a new curvature with sign $\sigma$, or it can assume a new curvature with sign $-\sigma$. New curvatures with the same sign are close to the initial curvature and are weighted by Boltzmann factors.
New curvatures of opposite sign must overcome an energy barrier $E_{max}$ and therefore are treated differently from
curvatures with the same sign. The outcome of the collision is chosen stochastically according to the probabilities represented in the
partition sum. Independent of this outcome, the resulting parasite deformation is only temporary, meaning that once the collision is solved the rod proceeds its migration with unit curvature. Progressively, the deformed region shifts towards the back of the rod. If the interval $I$ consists only of $\kappa=\pm 1$, we recover the binary model. When $I_c$ is empty, the rod performs a random reorientation, again as in the binary model. The inverse
temperature $\beta$ describes the importance of the energy terms for adhesion and bending versus an equal partitioning between
all possible states.

\subsection{Classification of the trajectories}

\begin{figure}
 \includegraphics{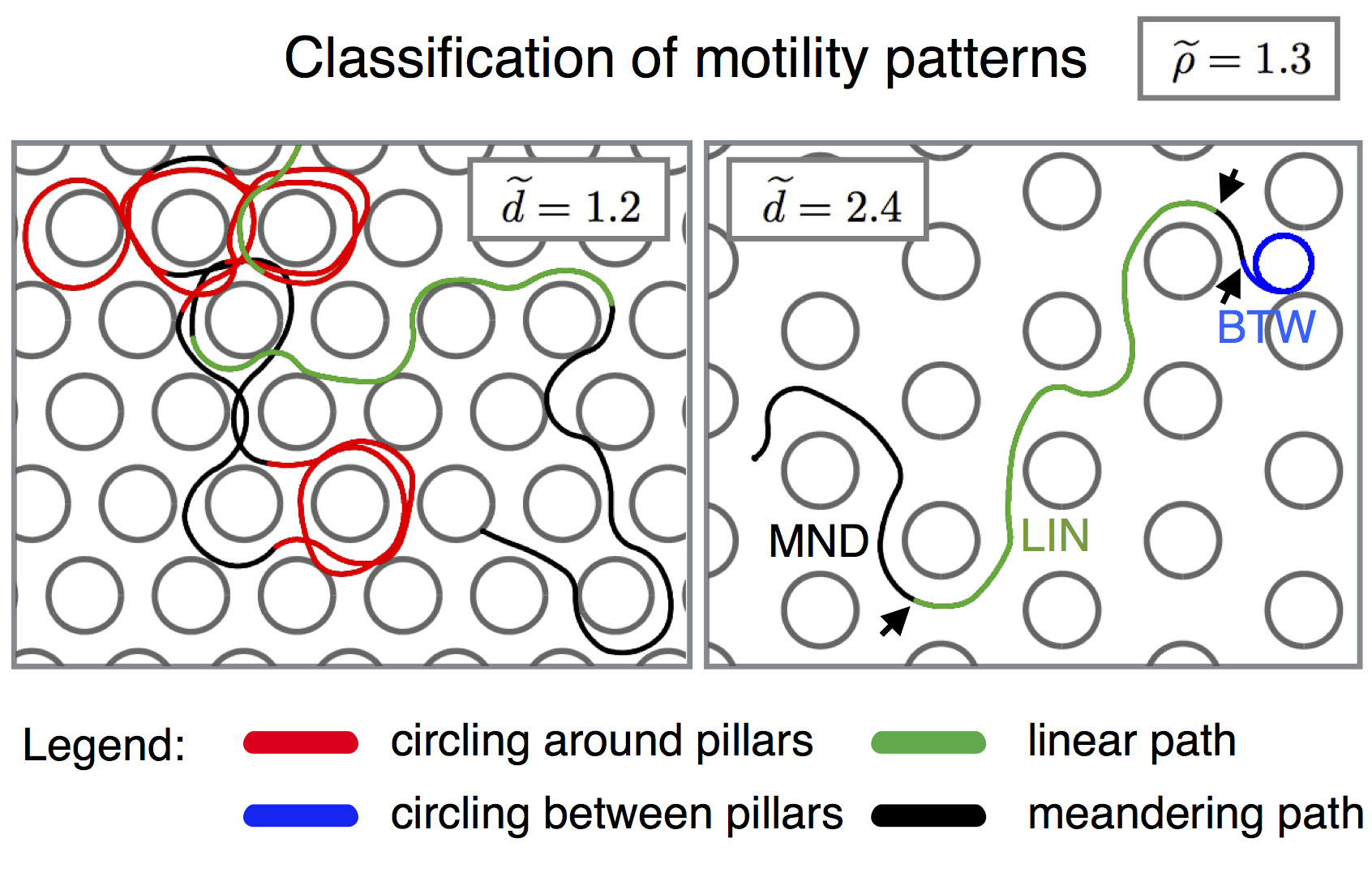}
 \caption{\label{fig:class_traj} Sample output of the algorithm classifying the motility patterns. For each trajectory we isolate paths circling around obstacles ($ARN$), between obstacles ($BTW$) and linear paths ($LIN$); the parts of the trajectory excluded from this categories are classified as meandering ($MND$). In the right box, the arrows identify the transitions from one pattern to another.}
\end{figure}

We analyse the trajectory the parasite describes under the different interaction rules by tracking its rear end $P$. Using a custom-made MATLAB routine we classify which proportion of the trajectory of $P$ is spent either circling around a pillar ($ARN$), circling between pillars ($BTW$), or moving through the lattice in a linear fashion ($LIN$). Portions of the trajectory not falling in any of these categories are classified as meandering ($MND$). Sample results of this classification are shown in fig.~\ref{fig:class_traj}. In the following we analyse the trajectories of the parasite for two representative values of $\widetilde{\rho}=0.7,1.3$ and a set of distances $\widetilde{d}\in[0.2,3]$ and compare the results with the predictions from the geometric diagram in fig.~\ref{fig:system_diagr}b. The details of our analysis of the trajectories are explained in appendix B. 

\subsection{Simulation results}

In our model, the parameters $\beta$ and $E_{max}$ describe the effects of adhesion and
bending. Because at the current stage it is not possible to estimate their values from
a microscopic model, we fix their values to $\beta=\log2$ and $E_{max}=1$ as a
representative choice for a wide set of parameters that lead to results consistent
with the experimental observations. In accordance with experimental observations, the typical parasite
length is set to $\widetilde{L}=2\pi / 3$. In the following, 
each parameter set is studied using $N=50$ independent simulations. The rod moves with constant unit speed and the trajectory is measured in terms of the non-dimensional arc length $s$. The rod is discretized in steps $ds=0.02\widetilde{L}$ and each simulation can last until the trajectory is $200\widetilde{L}$ long. For numerical reasons, a simulation is terminated when, at a random reorientation, the distance between $P$ and the obstacle is below $d_{min}=0.01\widetilde{L}$.

\begin{figure*}
 \includegraphics{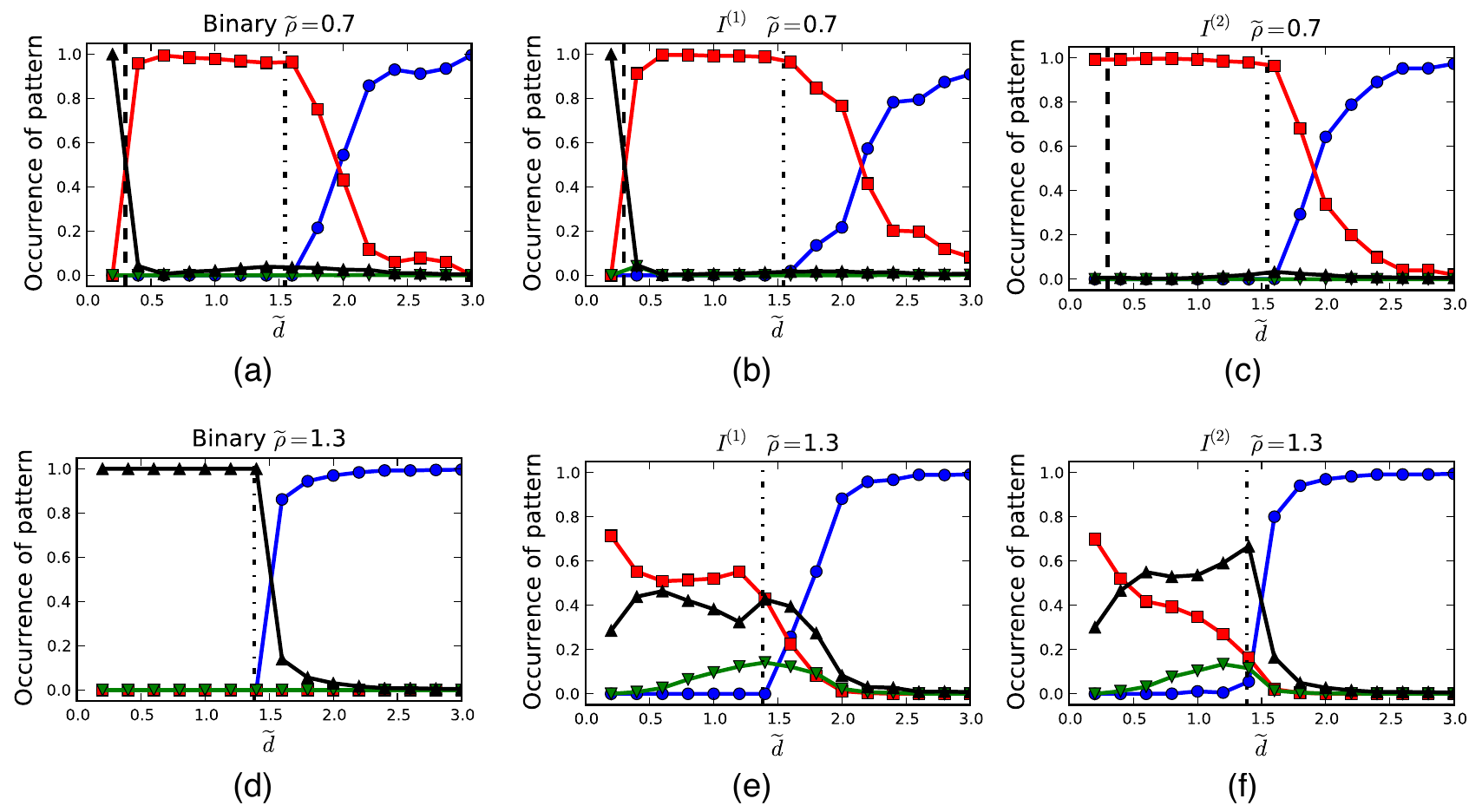}
 \caption{\label{fig:bin_st1_st2} Occurrence of motility patterns for stiff and flexible rods. The red (squares), blue (circles) and green (downward triangles) solid lines denote the $ARN$, $BTW$ and $LIN$ patterns respectively. The black solid line (upwards triangles) refers to $MND$ patterns. (a) For the binary model with $\widetilde{\rho}=0.7$ the distance dependence of the motility patterns is characterised by a transition from $MND$ to $ARN$ to $BTW$. The transition from $ARN$ to $BTW$ is conserved also when the rod is flexible (b,c), but the transition from $MND$ to $ARN$ is lost in (c), when the rod can increase its absolute curvature to avoid an obstacle. The dash-dotted and dashed lines represent the lines in fig.~\ref{fig:system_diagr}b separating region 1 from 2 and region 2 from 3 for $\widetilde{\rho}<1$. The transition from one motility pattern to another takes place as predicted by fig.~\ref{fig:system_diagr}b. (d) The binary model for $\widetilde{\rho}=1.3$ only allows for a transition from $MND$ to $BTW$. Flexible rods (b,c) can also circle around the obstacles and move linearly in the array. The dash-dotted line represents the line in fig.~\ref{fig:system_diagr}b separating region 1 from 4 for $\widetilde{\rho}>1$. It corresponds to the onset of $BTW$ patterns for both stiff and the flexible rods.}
\end{figure*}

We start with the condition $E_{tmb}=+\infty$, which means that the collision with an obstacle is solved by a random reorientation only when none of the possible deformations is able to free the rod. Results of the binary model, corresponding to the limiting case $\kappa=\pm 1$, are represented in figs.~\ref{fig:bin_st1_st2}a and \ref{fig:bin_st1_st2}d. For $\widetilde{\rho}=0.7$ (fig.~\ref{fig:bin_st1_st2}a) we observe, with increasing $\widetilde{d}$, a dominance of $MND$, then $ARN$ and finally $BTW$. The transition from $MND$ to $ARN$ takes place in correspondence with the dashed line, that represents the curve separating region 1 from region 2 in fig.~\ref{fig:system_diagr}b. Similarly, the transition from $ARN$ to $BTW$ begins at the dashed-dotted line, that represents the line separating region 2 and 3 in fig.~\ref{fig:system_diagr}b. 
For $\widetilde{\rho}=1.3$ we only observe a transition from $MND$ to $BTW$, because the obstacle radius is too large to allow for $ARN$. The transition coincides with the dashed-dotted line, that separates region 2 and 3 in fig.~\ref{fig:system_diagr}b.
Results from the binary model therefore agree very well with the predictions from the geometrical phase diagram. 

Allowing the rod to change curvature upon collision has the effect of increasing the occurrence of circling for $\widetilde{\rho}<1$ (figs.~\ref{fig:bin_st1_st2}b and \ref{fig:bin_st1_st2}c). For $\widetilde{\rho}>1$ the flexibility allows circling and linear patterns where in the binary case only meandering is possible (compare fig.~\ref{fig:bin_st1_st2}d with figs.~\ref{fig:bin_st1_st2}e and \ref{fig:bin_st1_st2}f). Figures \ref{fig:bin_st1_st2}b and \ref{fig:bin_st1_st2}e refer to a rod that can choose curvatures from the interval $I^{(1)}=[-1,1]$, while figs.~\ref{fig:bin_st1_st2}c and \ref{fig:bin_st1_st2}f correspond to a rod choosing from the wider interval $I^{(2)}=\left[-\left(\pi/\widetilde{L}+1/2\right),\left(\pi/\widetilde{L}+1/2\right)\right]$. Since in this last case the rod can increase its absolute curvature to solve a collision, $BTW$ competes with $ARN$ at lower $\widetilde{d}$ than with $I^{(1)}$, leading to increased $MND$ and $BTW$ pattern and a faster decrease of $ARN$.

\begin{figure}
 \includegraphics{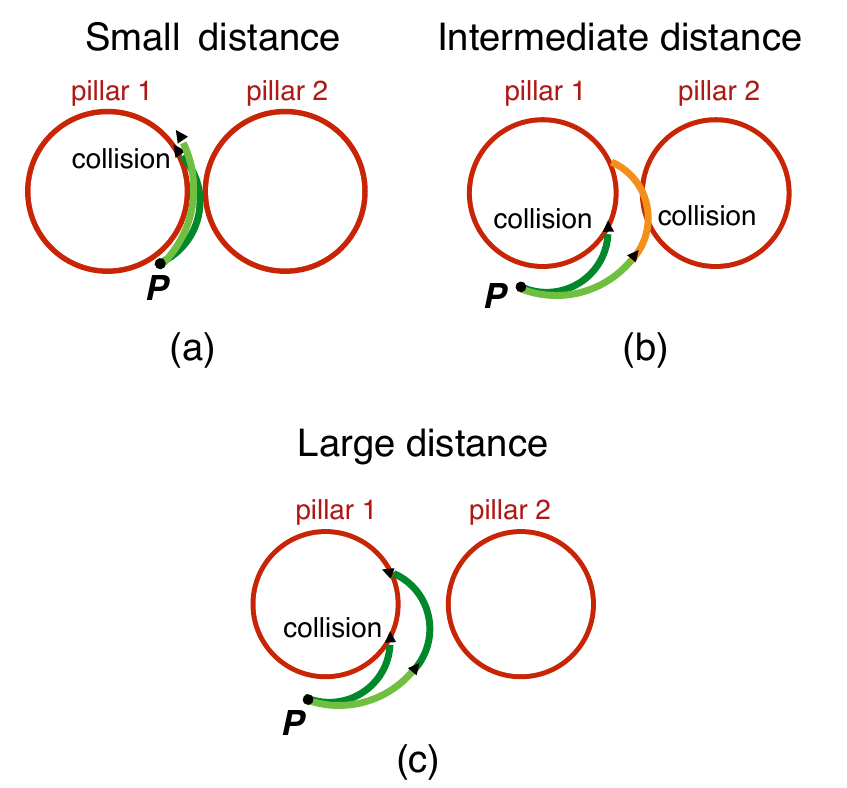}
 \caption{\label{fig:loc_min} Collisions of the parasite with the obstacles depending on the lattice spacing. (a) At small distances a collision causes the rod to deform based on pillar 1 and pillar 2. (b) Intermediate distances force the rod to interact separately with pillar 1 and pillar 2, while at large distances (c) the rod does not interact with pillar 2. Since intermediate distances correspond to an augmented collision frequency, random reorientations are also more frequent and circling around the obstacles is less likely than at small or large distances.}
\end{figure}

Figure \ref{fig:bin_st1_st2}e shows a dip in the $ARN$ curve. In fact, for pillar radii $\widetilde{\rho}>1$ circling can take place only if the rod is flexible, at the cost of continuous collisions with the obstacle and adjustments of the curvature of the rod. Three different situations can be isolated (fig.~\ref{fig:loc_min}). The distance $\widetilde{d}$ is very small (fig.~\ref{fig:loc_min}a): in this case the adjustment in the curvature of the rod depends on both \textit{pillar 1} and \textit{pillar 2}. For intermediate values of $\widetilde{d}$ (fig.~\ref{fig:loc_min}b) the change in curvature depends only on \textit{pillar 1}, but \textit{pillar 2} interferes with the trajectory of the rod after the curvature change. As a consequence, circling around \textit{pillar 1} is possible only at the cost of separate collisions with both \textit{pillar 2} and \textit{pillar 1}. For $\widetilde{d}$ large enough (fig.~\ref{fig:loc_min}c) \textit{pillar 2} does not interfere with the trajectory of the rod after the curvature change due to the collision with $\textit{pillar 1}$. Therefore, circling around \textit{pillar 1} does not require collisions with any other pillar.

Figure \ref{fig:bin_st1_st2}f does not show any dip in the $ARN$ curve; this is due to the interval $I^{(2)}$ being wider than $I^{(1)}$. The possibility of increasing the absolute curvature offers more possibilities to solve the collision at intermediate distances without randomly reorienting (fig.~\ref{fig:loc_min}). In addition, $BTW$ competes with $ARN$ at lower $\widetilde{d}$ than in the case of $I^{(1)}$, thereby favouring the decrease in the circling frequency at larger $\widetilde{d}$.

\begin{figure*}
 \includegraphics{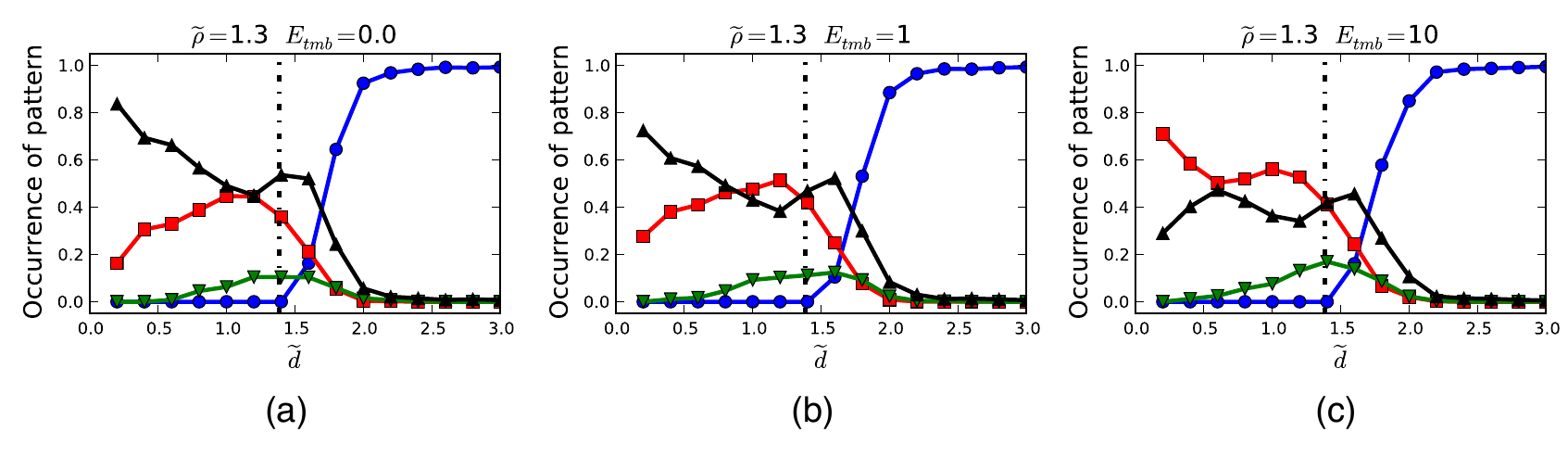}
 \caption{\label{fig:patt_tmb} Effect of a finite $E_{tmb}$ on the migration patterns of a flexible rod ($I^{(1)}$). The red (squares), blue (circles) and green (downward triangles) solid lines denote the $ARN$, $BTW$ and $LIN$ patterns respectively, the black solid line (upward triangles) refers to $MND$ patterns. The dash-dotted line represents the onset of $BTW$ as predicted by eq.~\eqref{eq:d_btw}. (a,b) Low values of $E_{tmb}$ allow the rod to solve a collision via a random reorientation even if the obstacle could be avoided via a change in curvature. This favours meandering patterns at low $\widetilde{d}$, when collisions with the obstacles are most frequent. (c) High values of $E_{tmb}$ restore the changes in the bending of the rod at low $\widetilde{d}$ (fig.~\ref{fig:bin_st1_st2}e, fig.~\ref{fig:loc_min}a).}
 \end{figure*}

We next consider the case of a finite $E_{tmb}$, which means that the parasite can solve collisions by tumbling even when they could be solved by a deformation. The effect of $E_{tmb}$ is relevant in conditions where collisions between the rod and the obstacles are frequent and the possible solutions limited, i.e.~for $\widetilde{\rho}>1$ and low $\widetilde{d}$ (fig.~\ref{fig:patt_tmb} for $I^{(1)}$). Low values of $E_{tmb}$ favour random reorientations, particularly at low distances. Consequently, meandering paths grow at low $\widetilde{d}$ and the dip in the $ARN$ disappears (figs.~\ref{fig:patt_tmb}a and \ref{fig:patt_tmb}b), while it is still present at higher $E_{tmb}$ (fig.~\ref{fig:patt_tmb}c).

\section{Model extensions}

\subsection{Anisotropic obstacle arrays}

\begin{figure*}
 \includegraphics{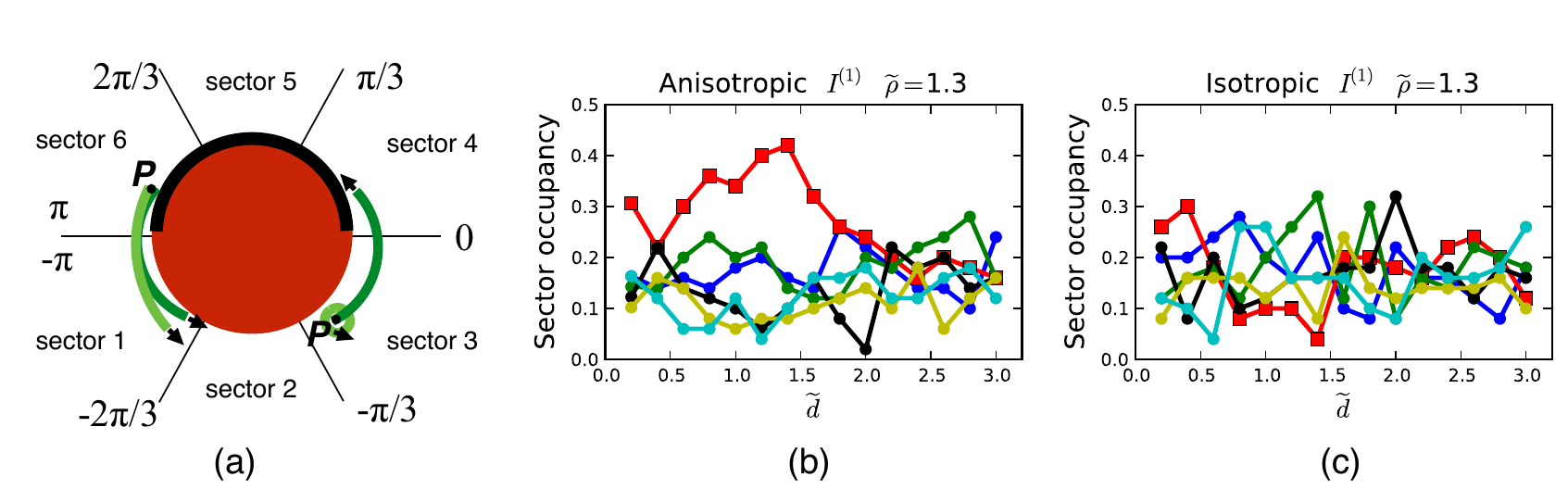}
 \caption{\label{fig:latt_coat} (a) An anisotropic obstacle array consists of pillars that have half of the contour (black) triggering a random reorientation of the rod, while the other half favours changes in the bending of the rod. We divide the lattice in sectors corresponding to the natural directions of the lattice in order to quantify the directionality imposed by the anisotropy on the trajectories. (b,c) Fraction of trajectories that end up in each of the sectors defined in (a). Sectors from 1 to 6 are plotted, in order, in blue, red, green, black, yellow and cyan. Sector 2 can be distinguished by the square markers. Anisotropic obstacle arrays (b) favour the occupancy of sector 2 (red line, square markers), whereas no sector is preferred in the isotropic case (c). The anisotropic obstacle array induces directionality because it renders less likely for the rod to move from a non-coated to a coated region than the other way around.}
\end{figure*}

We introduce an anisotropy in the obstacle array by assuming that half of the contour of the obstacle can trigger a random reorientation of the rod upon collision. Collisions occuring at the reorienting contour (black in fig.~\ref{fig:latt_coat}a) are assigned a tumbling energy $E_{tmb}=-20$, while we set $E_{tmb}=20$ otherwise. To analyse the effect of the anisotropy on the trajectories we divide the lattice in 6 sectors, each $\pi/3$ wide, and measure the fraction of the simulated trajectories that terminate in each of them. The sectors are centered at the starting point of each trajectory; in fig.~\ref{fig:latt_coat}a they are centred on the obstacle to emphasise the connection with the natural directions of the lattice.

The effect of the reorienting contour is pronounced for $\widetilde{\rho}>1$, that is when the rod can circle around an obstacle only via continuous collisions and curvature adjustments. Under the obstacle configuration of fig.~\ref{fig:latt_coat}a the rod is more likely to end up in \textit{sector 2}, represented by the red line (square markers) in fig.~\ref{fig:latt_coat}b (curvature interval $I^{(1)}$). In contrast, no sector stands out when we analyse the location of the end points in an isotropic obstacle array (fig.~\ref{fig:latt_coat}c).
The directionality in the migration induced by the anisotropy in the pillar array can be explained with the help of fig.~\ref{fig:latt_coat}a. A rod moving from the reorientation-triggering region to a normal region of the obstacle collides with it in the normal region and solves the interaction via a deformation. On the contrary, a rod that moves from the normal region to the reorientation-triggering region collides with the part of the pillar that forces a random reorientation. Since the centre of the random reorientation is the rear end $P$, the random reorientation takes place close to the normal region of the obstacle. Effectively, this feature renders less likely for the rod to move from a normal to a reorientation-triggering region than the other way around, thereby leading to the observed directionality.

\subsection{Heterogeneous obstacle arrays}

\begin{figure}
 \includegraphics{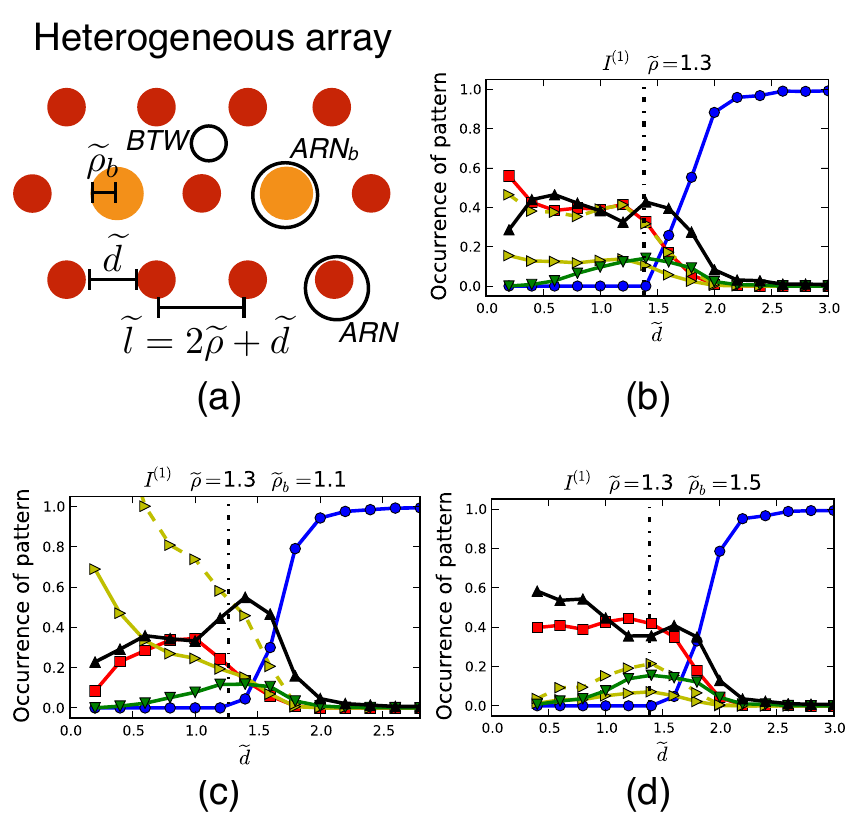}
 \caption{\label{fig:het_results} (a) Heterogeneous arrays include obstacles of two different radii, $\widetilde{\rho}$ and $\widetilde{\rho}_b$; $\widetilde{\rho}$ is three times more common than $\widetilde{\rho}_b$; $\widetilde{\rho}$ can be smaller or larger than $\widetilde{\rho}_b$. Stable motility patterns for a bent rod in the lattice are circling between obstacles ($BTW$) or circling around $\widetilde{\rho}$ or $\widetilde{\rho}_b$ ($ARN$ and $ARN_b$, respectively). (b,c,d) Occurrence of motility patterns for a flexible rod ($I^{(1)}$) in heterogeneous obstacle arrays. The red (squares), yellow (right triangles), blue (circles), green (downward triangles) and black (upward triangles) solid lines denote, in this order, $ARN$, $ARN_b$, $BTW$, $LIN$ and $MND$ patterns. The dashed yellow line (right triangles) corresponds to three times the frequency of $ARN_b$. (b) If $\widetilde{\rho}=\widetilde{\rho}_b$ the occurrence of $ARN$ reflects $\widetilde{\rho}$ being three times more common than $\widetilde{\rho}_b$. This correspondence is broken when $\widetilde{\rho}\neq\widetilde{\rho}_b$ in (c) and (d), because the rod associates more often with obstacles that represent a better geometrical fit.}
\end{figure}

Let us consider environments with obstacles of two different radii, as in fig.~\ref{fig:het_results}a. In the following we will refer to the outer (more common) pillar radius as $\widetilde{\rho}$, while the inner, less common pillar radius is labeled $\widetilde{\rho}_b$. We refer to the distance between two neighbouring outer pillars as $\widetilde{d}$.

Geometrical considerations analogous to those for the homogeneous obstacle array allow us to state that the rod moves steadily in the $(\widetilde{\rho},\widetilde{\rho}_b,\widetilde{d})$ lattice if it can circle between the pillars ($BTW$), around the outer pillar ($ARN$) or around the inner pillar ($ARN_b$) (fig.~\ref{fig:het_results}a). 
The pattern $ARN _b$ is possible for $\widetilde{\rho} _b<1$. The condition on $\widetilde{d}$ for $ARN _b$ to take place is $\widetilde{d}\geq 1-\widetilde{\rho}$. 
The pattern $ARN$ is possible for $\widetilde{\rho}<1$, while the corresponding conditions on $\widetilde{d}$ are:
\begin{align}
 &\widetilde{d}\geq 1-\widetilde{\rho}&\text{for}\,\,\widetilde{\rho} _b\leq\widetilde{\rho}\\
 &\widetilde{d}\geq1-2\widetilde{\rho}+\widetilde{\rho} _b&\text{for}\,\,\widetilde{\rho} _b>\widetilde{\rho}
\end{align}
Finally, $BTW$ can take place for 
\begin{align}
 &\widetilde{d}\geq \frac{1}{2}\left[ \sqrt{3}(1+\widetilde{\rho} _b) + \sqrt{ 4(1+\widetilde{\rho})^2 -(\widetilde{\rho} _b+1)^2 }\right] - 2\widetilde{\rho}  \nonumber \\
 &\hspace{5cm} \text{for}\,\,\widetilde{\rho} _b\leq\widetilde{\rho}\\
 &\widetilde{d}\geq \sqrt{3} - \widetilde{\rho}(2-\sqrt{3}) \hspace{1.9cm}\text{for}\,\,\widetilde{\rho} _b>\widetilde{\rho}
\end{align}

We now investigate how the trajectories of a flexible rod (curvature interval $I^{(1)}$) in a heterogeneous pillar array differ from those in a homogeneous array. In order to do that, we adapt the MATLAB routine for trajectory classification to distinguish circling around the two different kinds of pillars. As a control, we apply the routine to a flexible rod ($I^{(1)}$) with $E_{tmb}=+\infty$ in a homogeneous pillar array (fig.~\ref{fig:het_results}b). Coherently with the homogeneity of the array, the occurrence of circling around $\widetilde{\rho}_b$ is $1/3$ of the occurrence of circling around $\widetilde{\rho}$. The deviations between the dashed yellow line (right triangular markers), that is three times the occurrence of circling around $\widetilde{\rho}_b$ and the red line (square markers, circling around $\widetilde{\rho}$) are due to finite statistics effects.

Figures \ref{fig:het_results}c and \ref{fig:het_results}d show the analysis of the patterns for the two pillar pairs $(\widetilde{\rho},\widetilde{\rho}_b)=(1.3,1.1)$ and $(\widetilde{\rho},\widetilde{\rho}_b)=(1.3,1.5)$. In both cases, all the pillar radii exceed the radius of curvature of the rod, therefore making circling around any of the pillars possible only at the cost of continuous interactions and adjustments of the rod's curvature. Focusing on the circling patterns around each of the two available radii, we see that the rod associates preferably with the pillars that have the curvature closest to 1, avoiding the other option. In other words, the rod spends more time in the regions of the pillar array that represent a better geometrical fit. 

\subsection{3D obstacle arrays}

\begin{figure}
 \includegraphics{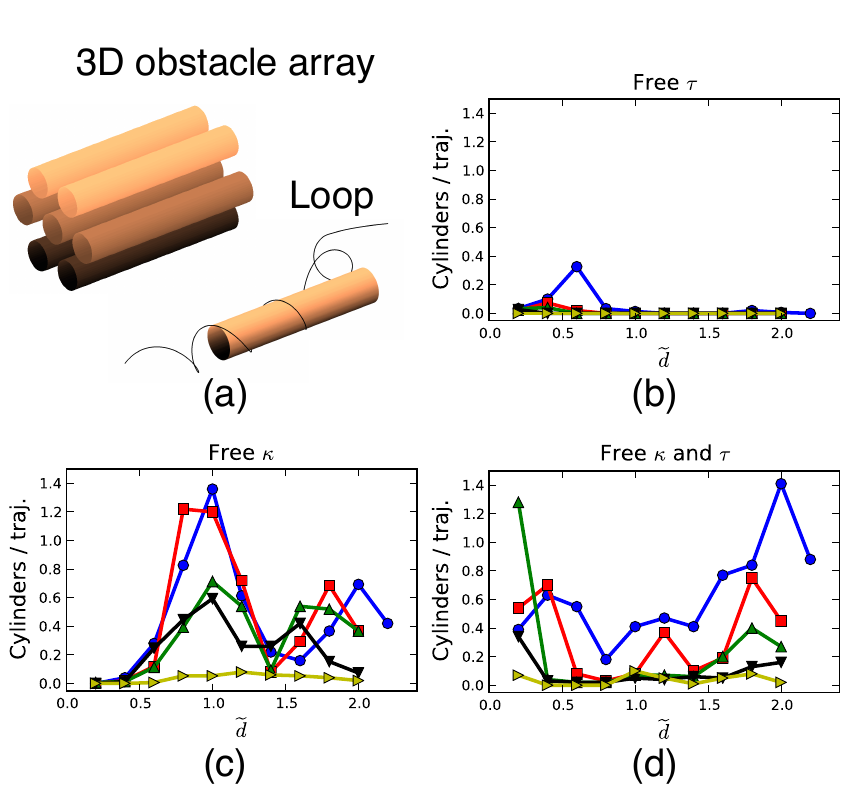}
 \caption{\label{fig:3d_results} (a) Three dimensional obstacles array consist of a identical cylinders arranged according to a hexagonal lattice. A bent and twisted rod that moves in the array on a helical path is deformed and deviated by the interaction with the obstacles and can as a result loop around the cylinders. (b,c,d) Mean number of cylinders per simulated trajectory involved in one or more looping events. The blue (circles), red (squares), green (upward triangles), black (downward triangles) and yellow (right triangles) curves refer, respectively, to $\widetilde{\rho}=0.6,0.7,0.8,0.9\,\text{and}\,1$. Looping is deterred by variations in the torsion of the rod (b), but is favoured at smaller radii for rods with a flexible curvature, as in (c) and (d).}
\end{figure}

In three dimensions, we have to consider not only bending, but also twisting as a fundamental deformation mode
of the parasite. We extend our model to three dimensions by adding a twist to the bent rod with the introduction of a non-dimensional torsion parameter $\widetilde{\tau}_0$. Motivated by the interaction between sporozoite and blood vessels,
as an obstacle array we now consider a hexagonal lattice of hard cylinders of radius $\widetilde{\rho}$ (fig.~\ref{fig:3d_results}a). The unperturbed trajectory of the rod is assumed to be a helix with unit curvature and torsion $\widetilde{\tau}_0$ so that, in terms of the non dimensional arc-length $s$, a trajectory around the $z$ axis is expressed as
\begin{equation}
\mathbf{h}_{\widetilde{\kappa},\widetilde{\tau}}(s) = 
\begin{pmatrix}
  \frac{1}{1+\widetilde{\tau}_0^2}\cos\left(s\sqrt{1+\widetilde{\tau}_0^2}\right) \\
  \frac{1}{1+\widetilde{\tau}_0^2}\sin\left(s\sqrt{1+\widetilde{\tau}_0^2}\right) \\
  \frac{s}{\sqrt{ 1+\widetilde{\tau}_0^{-2} }}
\end{pmatrix}
\end{equation}
The three dimensional equivalent of $ARN$ is a helical trajectory looping around a cylindrical obstacle. 

For simplicity we consider only positive curvatures and torsions. 
Let us consider a set of curvatures $I=[\widetilde{\kappa}_{min},\widetilde{\kappa}_{max}]$ and a set of torsions $J=[\widetilde{\tau}_{min},\widetilde{\tau}_{max}]$ that the rod can assume in order to avoid a cylinder upon collision. To each configuration of the rod we assign an energy $E$:
\begin{align}
 E = & \int_0^{\widetilde{L}} ds\,[ E_{\kappa}(\kappa(s))+E_{\tau}(\tau(s)) ] = \nonumber \\ 
 = & \int_0^{\widetilde{L}} ds\,[ (\kappa(s)-1)^2 + \alpha(\tau(s)-\widetilde{\tau}_0)^2 ]
\end{align}
where $\alpha$ is a coefficient defining the relative contribution of the curvature and torsion energy terms. Similarly to the two dimensional case (eq.~\eqref{eq:Zc}) we define the partition function
\begin{align}
 Z(\beta) = & \sum_{\kappa\in I}\sum_{\tau\in J}e^{-\beta[E_{\kappa}(\kappa(s))+E_{\tau}(\tau(s))]} = \nonumber \\
 = & \sum_{\kappa\in I_c}\sum_{\tau\in J_c}e^{-\beta[E_{\kappa}(\kappa(s))+E_{\tau}(\tau(s))]}\enspace.
\end{align}
In order to avoid overly expensive computations,
we moreover use the zero temperature limit ($\beta \to \infty$), which constrains the system to its minimal energy
states. The pair $(\widetilde{\kappa},\widetilde{\tau})$ that solves a collision is the one minimising $E$ in $I_c\times J_c$.

We choose $\widetilde{L}=2\pi/3$, $\alpha=1$ and $\widetilde{\tau}_0=0.5$. Each parameter set is supported by $N=150$ independent simulations when the rod is flexible in the curvature or in the torsion only, by $N=100$ when both curvature and torsion can vary. The rod is discretised in steps $ds=0.02\widetilde{L}$ while each simulation can last until the trajectory is $200\widetilde{L}$ long. A simulation is terminated if no $(\widetilde{\kappa},\widetilde{\tau})$ configuration allows the rod to solve the collision. The intervals for curvature and torsion are $\left[0,2\pi/\widetilde{L}\right]$ and $\left[0,\sqrt{ (2\pi/\widetilde{L})^2-1}\right]$ respectively.

We compare the trajectories of rods that are allowed to solve collisions by changing only their curvature, only their torsion, or both. Trajectories are characterised by the number of cylindrical obstacles involved in at least one looping event (fig.~\ref{fig:3d_results}a). We find that looping around the obstacles is favoured when the rod is allowed to change its curvature upon collision, but is suppressed when only the torsion is allowed to change (figs.~\ref{fig:3d_results}b and \ref{fig:3d_results}c). When both curvature and torsion are variable, the rod can loop around the obstacles also in very crowded environments (fig.~\ref{fig:3d_results}d).

\section{Discussion}

Fast and effective sporozoite migration in the host skin is essential to ensure successful transmission of malaria \cite{Sidjanski1997,Vanderberg2004,Amino2006}. Starting from the striking observation that the sporozoite stage
is the only phase of the life cycle of the malaria parasite in which the cells are crescent-shaped \cite{Vanderberg1974},
we have studied the consequences of this peculiar geometry for sporozoite motion in structured
environments. Existing studies addressing sporozoite motility have mainly focused on the molecular mechanisms underlying the propulsion machinery \cite{Kappe2004,Montagna2012}. Here we take a more macroscopic viewpoint and build on
previous work that has investigated experimentally the migration of sporozoites in pillar arrays \cite{Hellmann2011}.
No theoretical study of the mechanical interplay between the parasite and the environment has, to our knowledge, been presented so far. In this paper we have used geometrical arguments and stochastic computer simulations to analyze sporozoites as self-propelled and curved rods. We have focused on migration in arrays with circular or cylindrical obstacles because of our focus on geometrical determinants.

For arrays of identical circular obstacles, the system is most interesting when the obstacles have radii larger than one, so that the bent rod, that has unit curvature, can associate with the obstacles only at the cost of continuous collisions. In this case, the migration patterns of flexible parasites are much richer than those of stiff parasites. Flexible parasites can associate to the obstacles and also move linearly in the lattice where, under the same conditions, stiff parasites can only collide and reorient, resulting in meandering patterns (figs.~\ref{fig:bin_st1_st2}d, \ref{fig:bin_st1_st2}e and \ref{fig:bin_st1_st2}f). A flexible parasite can solve its collisions via a change in curvature, therefore reducing the frequency of random reorientations and describing a trajectory that in general is more stable and smoother.

Flexible parasites are sensitive to anisotropies in the obstacle array (fig.~\ref{fig:latt_coat}a), that trigger directional movement (fig.~\ref{fig:latt_coat}b). This finding provides an interesting experimental perspective. In fact, selectively coated obstacles offer a simple coupling between mechanical and chemical properties in sporozoite motility. Assays of this kind are in principle experimentally realisable \cite{Richter2013}, provided a suitable inhibitor of adhesion to the substrate and the pillars is used. For example, polyethylene glycol (PEG) has been used previously to prevent sporozoite adhesion \cite{Perschmann2011}. Such a system would contribute to the validation and improvement of the model, in addition to further help the understanding of sporozoite motility in controlled environments and in the presence of an elementary anisotropy. Indeed, one could envision experimental setups of intermediate complexity, for example arrays of soft elastic pillars, that could give information on the forces involved in collisions. Insights from such systems would then help to quantitatively analyse sporozoite motility \textit{in vivo}. 

We have further used our model to analyse the effect of obstacles of different sizes (fig.~\ref{fig:het_results}a) on the trajectories of the rod. Our results show how the sporozoite statistically associates with preference to the obstacles that provide a better fit to its shape (fig.~\ref{fig:het_results}). It is tempting to interpret this result \textit{in vivo}, when the sporozoite glides through the dermis of the host aiming at a blood capillary. The ability to spend more time around structures of matching size could help the parasite in the search for a blood capillary to invade. Surprisingly, the typical radius of a blood capillary is about $8-10\ \mu m$, very close to the typical radius of curvature of sporozoites \cite{Vanderberg1974}, that are also observed to loop around blood vessels before invading them \cite{Amino2006}. 

As a last step, we have extended the model to account for parasite migration in three dimensional obstacle arrays (fig.~\ref{fig:3d_results}a). In this case, the parasite is both bent and twisted and moves, if unperturbed, on a helical trajectory. Collisions with the cylindrical obstacles deviate the rod and allow for the trajectory to loop around them (fig.~\ref{fig:3d_results}a). This feature is intriguing, in that it closely resembles the circling of sporozoites around blood vessels \textit{in vivo} \cite{Amino2006}. Our findings reveal that flexibility in the curvature of the parasite is essential for looping to take place. Torsional flexibility alone does not allow for looping, but combined with curvature variations boosts looping in very packed environments (fig.~\ref{fig:3d_results}). We speculate that sporozoite motility \textit{in vivo} and the association to blood capillaries is greatly favoured by the ability of the parasite to adapt its curvature and torsion based on the structure of the surrounding environment. New techniques to custom design three dimensional micro-environments \cite{Connell2013} represent a high potential for further experimental studies of sporozoite migration.

To conclude, we have shown that sporozoites migrating in complex environments can be efficiently modelled as self-propelled bent (and, in three dimensions, twisted) rods. The flexibility in the bending of the rods facilitates migration in obstacle arrays that do not represent a good geometrical fit. Our focus on cell shape and on the details of the interaction with the obstacles shows that much of the complexity observed in experiments studying sporozoite migration \cite{Hellmann2011} can indeed stem from the geometrical and mechanical properties of the system. Using three-dimensional cylindrical obstacles we have shown that trajectories can loop around them, suggesting a prominent role of the geometrical properties of sporozoites also during migration in the skin of the host, that is successful if it terminates with the finding and invasion of a blood capillary. Our work provides
a theoretical framework to analyze future experiments in a quantitative manner.

We finally note that our modelling approach, focusing on basic geometrical and mechanical properties, is not limited to the study of sporozoite motility. In fact, it could be used for the study of other motile cells that undergo temporary shape changes during gliding, for example other apicomplexa such as \textit{Toxoplasma gondii} \cite{Hakansson1999} or gliding bacteria \cite{McBride2001}.

\begin{acknowledgments}
We thank Janina Hellmann, M. Muthinja, Lucas Sch\"{u}tz, Noa Dahan, Leandro Lemgruber and Rogerio Amino for contributing images and Miriam Ester for mosquito production and animal handling. AB was funded by the Heidelberg Graduate School of Mathematical and Computational Methods for the Sciences.
FF was funded by the Chica and Heinz Schaller Foundation, the Human Frontier Science Program Organisation (RGY0071/2011) and the EU FP7 Network of Excellence EVIMalaR. FF and USS are members of the Heidelberg cluster of excellence CellNetworks and of the collaborative research center SFB 1129 on integrative analysis of pathogen replication and spread. 
\end{acknowledgments}

\appendix
\section{Details on the experiments \label{sec:det_exp}}

\subsection{Ethic Statement.} 
All animal experiments were performed concerning FELASA category B and GV-SOLAS standard guidelines. Animal experiments were approved by German authorities (Regierungspr\"{a}sidium Karlsruhe, Germany), \S 8 Abs.~1 Tierschutzgesetz (TierSchG). 

\subsection{Plasmodium sporozoite generation and imaging.} 
Fluorescent Plasmodium \textit{berghei} (strain NK65) sporozoites were produced in \textit{Anopheles stephensi} mosquitoes and isolated as described in Ref.~\cite{Hellmann2010}. Sporozoites were harvested between 17-21 days after mosquito infection. Imaging of motile sporozoites was performed on an inverted Axiovert 200M Zeiss microscope using the GFP filterset 37 (450/510) at room temperature. Images were collected with a Zeiss Axiocam HRM at 1 Hz using Axiovision 4.6 software and a 10x Apoplan objective lens (NA=0.25). DIC (differential interference contrast) images of the substrate were taken before and after the movie sequence to generate merged files of the PDMS pillars with the time-lapse series (or projected trajectories) of motile sporozoites. Generally, a time lapse series was recorded that comprised one image per second for a total of 300 frames. The parasites were either tracked manually using the ImageJ manual tracking plugin or with ToAST \cite{Hegge2009}. Images were compiled with ImageJ. 

\subsection{Production of micro-pillar substrates.} 
We used the micro-pillar arrays described in Ref.~\cite{Hellmann2011} plus a new one exhibiting pillars of different diameters.
Briefly, a master mask was designed using a custom-made program. This was converted with the mask-writer software DWL-66 (Heidelberg Instruments). On the mask, the individual pillars were arranged in hexagonal patterns to minimise deviations in the desired distance between pillars in any direction. The master mask was produced by ML\&C Jena. The PDMS micro-pillar substrates were then made by photolithographic and replicate moulding techniques as described in Ref.~\cite{Roos2005}. To improve the wetting of the hydrophobic pillar substrates they were first treated chemically with Extran an alkaline solution (Merck, Germany) in a 1:10 dilution in $H_2O$ for 20 minutes on a shaker. Afterwards the pillar arrays were washed three times for 10 minutes in distilled water. A silicon flexiPERM chamber (GreinerBioOne) was used to record sporozoites, which surrounds the micro-pillar substrate and thus maintains a stable environment essentially without flow. For the motility studies on pillar substrates the sporozoites in motility medium (RPMI - Roswell Park Memorial Institute - including 3\% bovine serum albumin) were added to the chamber surrounding one pillar field.

\subsection{Curvature measurements}
We have measured the curvatures of fig.~\ref{fig:hist_curv} starting from images of fluorescent sporozoites on a flat substrate. We have used MATLAB routines to extract the morphological skeleton of each cell. We have then computed the radius of the circumcircle of all the triangles that can be built using the end points of the skeleton as two of the vertexes. The curvature of the parasite is defined as the inverse of the average of the circumcircle radii.

\section{Classification of motility patterns \label{sec:class_mot_patt}}
\subsection{Two dimensional obstacle arrays}
To analyse the motion of a bent rod in an obstacle array we study the positions of its rear end $P$. The trajectory of $P$ is continuous everywhere, except when a random reorientation takes place. In this case, the bent rod rounds up with centre $P$ and uniformly samples an exit angle, from which it resumes migration with unit curvature and a probability $1/2$ of moving either clockwise or counterclockwise. The maximum length required for the point $P$ to lose memory of the rounding-up is $\widetilde{L}$.

The analysis of a trajectory $T$ begins with a check for random reorientations and the exclusion of the $\widetilde{L}$ long paths following them. Let $n_{r}$ be the number of random reorientations in $T$, so that there are at most $n_{r}+1$ continuous paths $T_i$ left to be analysed independently. Within each $T_i$, we distinguish between linear ($LNR$) paths, paths circling between pillars ($BTW$), around a pillar ($ARN$) as well as meandering paths ($MND$). In the case of arrays with obstacles of two different radii we distinguish circling around each of the two kinds ($ARN$ and $ARN_b$).

We define minimum lengths for the continuous paths $T_i$ to be possibly classified as $LNR$, $BTW$ or $ARN$. Paths longer than $L_{LNR}=8\widetilde{L}$ are considered for $LNR$ classification. Paths longer than $L_{BTW}=2\pi/\widetilde{\kappa}_0$ are considered for $BTW$ classification. Paths longer than $L_{ARN}=3\pi\max (1/\widetilde{\kappa}_0,\rho')$ are considered for $ARN$ classification. For homogeneous arrays, $\rho'=\widetilde{\rho}$, while for heterogeneous arrays $\rho'=(\widetilde{\rho}+\widetilde{\rho}_b)/2$. The default curvature of the rod $\widetilde{\kappa}_0$ is $1$ with our choice of non dimensional quantities, but is explicitly written here for clarity purposes. Paths $T_i$ shorter than $\min(L_{LNR},L_{BTW},L_{ARN})$ are classified as $MND$. Distinction between $LNR$, $BTW$ and $ARN$ paths takes place independently. Possible overlaps are taken care of as a final step.

\subsection{Circling between pillars} 
We compute the centre of mass $CM$ of each continuous piece of length $L_{BTW}$ of a trajectory $T_i$. The segment is classified as $BTW$ if $CM$ is outside the pillars, its distance from the closest pillar is larger than $\widetilde{\kappa}_0^{-1}$ and $\langle |T_i-CM|^2 \rangle$ is below a user-defined threshold that we set to $t_{BTW}=0.1/\widetilde{\kappa}_0$.

\subsection{Circling around pillars} 
For each continuous piece of length $L_{ARN}$ of a trajectory $T_i$ we compute the centre of mass $CM$. The segment is classified as $ARN$ in two cases. (a) $CM$ is outside the pillars, its distance from the closest pillar is below $\widetilde{\kappa}_0^{-1}-2\widetilde{\rho}^*$ and $\langle |T_i-CM|^2 \rangle$ is below a user-defined threshold $t_{ARN}$. (b) $CM$ is inside a pillar and $\langle |T_i-CM|^2 \rangle<t_{ARN}$. In both cases, we choose $t_{ARN}=0.3/\widetilde{\kappa}_0$. For homogeneous pillar arrays $\widetilde{\rho}^*=\widetilde{\rho}$, while for heterogeneous arrays $\widetilde{\rho}^*$ is the radius of the pillar that the parasite is circling around.

\subsection{Linear} 
For each continuous piece of length $L_{LNR}$ of a trajectory $T_i$ we compute the elongation as the ratio between the end-to-end distance and $L_{LNR}$. The segment is classified as $LNR$ if the elongation is above the user-defined threshold $t_{LNR}=0.8$.

All the regions that have not been classified up to now are classified as $MND$. We take care of the regions that have been classified as more than one pattern by ranking $ARN$ before $BTW$, which in turns ranks before $LNR$.

%\bibliography{refs2Dmodel.bib,refs2Dmodel_2.bib}


\end{document}